\documentclass{article}
\usepackage{arxiv}
\usepackage[utf8]{inputenc} %
\usepackage[T1]{fontenc}    %
\usepackage{hyperref}       %
\usepackage{url}            %
\usepackage{booktabs}       %
\usepackage{tabularx,ragged2e,booktabs,subcaption}
\newcolumntype{Y}{>{\RaggedRight\arraybackslash}X}
\usepackage{amsfonts}       %
\usepackage{nicefrac}       %
\usepackage{microtype}      %
\usepackage{lipsum}		%
\usepackage{graphicx}
\usepackage[square,numbers,sort&compress]{natbib}
\usepackage{doi}
\usepackage[misc]{ifsym}

\usepackage[inline]{enumitem}
\usepackage{subcaption}
\usepackage[export]{adjustbox}
\usepackage{amsmath}
\usepackage{color}
\definecolor{myblue}{RGB}{13, 71, 161} 
\definecolor{mygreen}{RGB}{115, 140, 84} 
\definecolor{myred}{RGB}{216, 28, 56}

\title{Taiji Data Challenge for Exploring Gravitational Wave Universe}
\date{\today}	%
\author{
        Zhixiang Ren\textsuperscript{1,2}\thanks{contributed equally}, 
        Tianyu Zhao\textsuperscript{1,3,2}\footnotemark[1], 
        Zhoujian Cao\textsuperscript{1,3,4}\footnotemark[2], 
        Zong-Kuan Guo\textsuperscript{1,5,6,4},  
        \\ \textbf{Wen-Biao Han\textsuperscript{1,4,7,8,9,10}, 
        Hong-Bo Jin\textsuperscript{1,11,8}, 
        Yue-Liang Wu\textsuperscript{1,5,4,9}\thanks{Corresponding author(s); \href{mailto:zjcao@amt.ac.cn}{\texttt{zjcao@amt.ac.cn}}; \href{mailto:ylwu@itp.ac.cn}{\texttt{ylwu@itp.ac.cn}}}
        }
        \bigskip
        \\ \textsuperscript{1} Taiji Laboratory for Gravitational Wave Universe, 
        \\ University of Chinese Academy of Science (UCAS), Beijing 100049, China
	    \\ \textsuperscript{2} PengCheng Laboratory, Shenzhen 518055, China
	    \\ \textsuperscript{3} Department of Astronomy,          Beijing Normal University, Beijing 100875, China 
        \\ \textsuperscript{4} School of Fundamental Physics and Mathematical Sciences, Hangzhou Institute for Advanced Study,
        \\ UCAS, Hangzhou 310024, China
        \\\textsuperscript{5} CAS Key Laboratory of Theoretical Physics, Institute of Theoretical Physics,
        \\ Chinese Academy of Sciences, Beijing 100190, China
	    \\ \textsuperscript{6} School of Physical Sciences,      UCAS, Beijing 100049, China
        \\ \textsuperscript{7} Shanghai Astronomical Observatory, Chinese Academy of Sciences, Shanghai 200030, China
        \\ \textsuperscript{8} School of Astronomy and Space Science, UCAS, Beijing 100049, China
        \\ \textsuperscript{9} International Centre for Theoretical Physics Asia-Pacific (ICTP-AP, UNESCO), UCAS, Beijing 100190, China
        \\ \textsuperscript{10} Shanghai Frontiers Science Center for Gravitational Wave Detection, Shanghai 200240, China
        \\ \textsuperscript{11} Key Laboratory of Computational Astrophysics, 
        \\ National Astronomical Observatories, Beijing 100101, China
 }

\renewcommand{\shorttitle}{Taiji Data Challenge}

\begin{document}
\maketitle

\begin{abstract}
	The direct observation of gravitational waves (GWs) opens a new window for exploring new physics from quanta to cosmos and provides a new tool for probing the evolution of universe.
	GWs detection in space covers a broad spectrum ranging over more than four orders of magnitude and enables us to study rich physical and astronomical phenomena.
	Taiji is a proposed space-based GW detection mission that will be launched in the 2030s.
	Taiji will be exposed to numerous overlapping and persistent GW signals buried in the foreground and background, posing various data analysis challenges.
	In order to empower potential scientific discoveries, the Mock Laser Interferometer Space Antenna (LISA) Data Challenge and the LISA Data Challenge (LDC) were developed.
	While LDC provides a baseline framework, the first LDC needs to be updated with more realistic simulations and adjusted detector responses for Taiji's constellation.
	In this paper, we review the scientific objectives and the roadmap for Taiji, as well as the technical difficulties in data analysis and the data generation strategy, and present the associated data challenges.
	In contrast to LDC, we utilize second-order Keplerian orbit and second-generation time delay interferometry techniques.
	Additionally, we employ a new model for the extreme-mass-ratio inspiral waveform and stochastic GW background spectrum, which enables us to test general relativity and measure the non-Gaussianity of curvature perturbations.
		Furthermore, we present a comprehensive showcase of parameter estimation using a toy dataset. This showcase not only demonstrates the scientific potential of the Taiji Data Challenge but also serves to validate the effectiveness of the pipeline.
	As the first data challenge for Taiji, we aim to build an open ground for data analysis related to Taiji sources and sciences.
	More details can be found on the official website \url{http://taiji-tdc.ictp-ap.org}.
\end{abstract}


\section{Introduction}
\label{sec1}
Ground-based GW detectors have made remarkable achievements since the first detection of GW150914 \cite{abbott_observation_2016},
nearly 100 GW events have been observed so far \cite{the_ligo_scientific_collaboration_gwtc-3_2021}.
The discovery of GWs not only provides a more fundamental test on Einstein's theory of general relativity (GR) but also leads us to study deeply on the nature of gravity and spacetime.
In particular, GWs detection in space enables us to probe the formation and structure of black holes from the early Universe through the peak of star formation era and to understand how does the intermediate mass seed black hole formed in the early universe and how the seed black hole grows into the large or extreme-large black hole \cite{gonzalez_2013}.
It will also be helpful to explore whether the dark matter could form into the black hole or primordial black hole could be the candidate of dark matter.

Due to the limitation by seismic noise and arm length, ground-based detectors are not sensitive at frequencies lower than 10 Hz \cite{Matichard_2015}.
By moving the detector to space, we can eliminate the seismic and gravity-gradient noise and open up the frequency band of the GW within $[10^{-4},0.1]\mathrm{Hz}$.
The first planned space-based GW detection project is the laser interferometer space antenna (LISA) \cite{amaro-seoane_laser_2017}, proposed by the European Space Agency (ESA) and the National Aeronautics and Space Administration (NASA), which will be launched in around 2037 \footnote{\url{https://sci.esa.int/web/lisa}}.
Furthermore, LISA has laid a solid foundation for space-borne GW detection, including scientific objectives, data analysis algorithms, and instrumental techniques.

During the early 2000s, Chinese scientists were also interested in space-borne GW detection programs and considered developing independent missions.
The feasibility of space-based GW detection was studied by the Chinese Academy of Sciences (CAS) in 2008, followed by the initial mission proposal presented in 2011 \cite{gong_scientific_2011}.
A half-million-kilometer arm length and a noise budget that is a hundred times more sensitive than LISA at 0.1 Hz made this instrument particularly sensitive to intermediate mass black hole mergers \cite{luoTaijiProgramConcise2021}.
In 2012, the ﬁrst mission descoping attempt was reported to prefer a constellation of three satellites separated each by about 3 million kilometers, coordinating both technical and scientiﬁc goals, and the proposed three-step road map was ﬁrst presented in which the launch time for the Chinese space GW detector was expected to be in the 2030s \cite{wu_2012,luoTaijiProgramConcise2021}.
The CAS built prototypes that were used to conduct proof-of-principle experiments \cite{li_laser_2012,liu_evaluation_2014,dong_methodological_2014,li_path-length_2015}.
Similar to LISA, Taiji consists of three spacecraft (SC), and each SC follows a heliocentric orbit.
The three SC form an approximately 3 million-kilometer-long equilateral triangle.
There is a 20-degree trail between the center of mass of the constellation and the Earth.
The constellation is roughly $1 \mathrm{AU}$ distant from the Sun. This caused a slight eccentricity in the orbits of the three SC.
With the stable constellation, Taiji's sensitive frequency band is $[10^{-4},0.1]
	\mathrm{Hz}$ \cite{luoTaijiProgramConcise2021}.
A three-step road map of the Taiji program is:
\begin {enumerate*} [label=\itshape\arabic*\upshape)]
\item launch a satellite Taiji-1 for individual technique validation,
\item launch two satellites Taiji-2 for almost all techniques test, and
\item launch complete Taiji three satellites \cite{hu_taiji_2017,luoBriefAnalysisTaiji2020}.
\end {enumerate*}
The Taiji program, with its three-step road map, has received priority support from the CAS's strategic priority research program since 2016 for its pre-experimental study.
Some key technologies have been developed further in the studies \cite{hu_taiji_2017,luo_possible_2017,luo_recent_2018,liu_principle_2018,liu_development_2018,deng_high-efficiency_2018,wang_zhi_preliminary_2018}.
Currently, the first phase of the Taiji-1 on-orbit experiment has been successfully completed \cite{the_taiji_scientific_collaboration_chinas_2021,thetaijiscientificcollaborationTaijiProgramSpace2021}.

In this paper, we present the first Taiji Data Challenge (TDC).
Specifically, we review the scientific objectives and technical challenges during the data processing procedure.
Then, we visualize all TDC datasets and explain their scientific potential and technical obstacles.
We implement time-domain time delay interferometry (TDI) 2.0 for detector response calculation, which makes our data more realistic compared to LDC1's original datasets.
As the main component of our data challenge, we generate datasets that contain only one source and also datasets that contain multiple sources.
We further generate an extreme-mass-ratio inspiral (EMRI) waveform constructed under the general parameterized metric, which allows us to test GR and investigate the non-Gaussianity of curvature perturbations by observing the GW background it induces.

The paper is organized as follows. In Section \ref{sec:SO}, we introduce the scientific objectives and technical challenge of Taiji mission. In Section \ref{sec:method}, we demonstrate how to simulate Taiji instrumental noise, detector response, and GW waveform respectively.  In Section \ref{sec:data}, we show the properties of each dataset. %
The last section contains our discussions about future work and features of our TDC website.

\section{Scientific objective and technical challenge of Taiji mission}
\label{sec:SO}
Taiji detectors will receive a large amount of GW signals from different sources in the target frequency band of $[10^{-4},0.1]\mathrm{Hz}$ \cite{hu_taiji_2017}.
The observatory will measure signals from a wide range of different sources relevant to the astrophysical mechanism of the formation of black holes and galaxies, the test of GR, and the study of cosmology, including massive black holes binaries (MBHBs) at all redshifts; EMRIs; the inspiral of stellar-origin binary black holes (SOBBHs); galactic binaries (GB); verification galactic binaries (VGBs), and various stochastic GW background (SGWB) sources \cite{luoTaijiProgramConcise2021}.

The mission's primary objective is to determine how and when massive black holes have formed and grown over cosmic time.
For the purpose of reconstructing their evolution, the study will explore almost all of the mass-redshift parameter space that is relevant \cite{the_taiji_scientific_collaboration_chinas_2021}.
We will investigate the distribution of MBHB's spins and masses in order to distinguish between different formation mechanisms based on the GW signal from MBHBs \cite{PhysRevD.93.024003}.
Additionally, the mission will analyze the signals from GBs and provide information on stellar evolution.
Tens of millions of binaries in our galaxy produce stochastic foregrounds, which we can estimate by observing GBs \cite{zhang_resolving_2021}.
It is possible to constrain the evolutionary pathways of compact binaries by examining the characteristics of the population, such as the number density of sources as a function of frequency.
It is expected that Taiji will provide exceptionally strong tests of GR predictions by observing highly relativistic MBHBs and EMRIs \cite{amaro-seoane_laser_2017}.
During the merger phase of a binary black hole system, BHs travel at nearly the speed of light and interact strongly with each other, which allows the study of the full nonlinear dynamics of gravity \cite{sesana_prospects_2016}.
By observing the signal of the EMRI system, the mass, spin, and quadrupole moment of the central massive black hole will be measured making testing its level of Kerr-ness and proving the no-hair hypothesis possible \cite{xin_gravitational_2019}.
Finally, a space-based GW detector will explore new physics and cosmology and seek out previously unknown GW sources.
The scientific objectives of the Taiji mission are listed in Table \ref{tab:SO}.

\begin{table}[h!]
	\centering
	\caption{Scientific objectives of Taiji mission}
	\label{tab:SO}
	\begin{tabularx}{\textwidth}{@{} lYl @{}}
		\toprule
		\textbf{No.} & \textbf{Scientific objective}                             \\
		\midrule
		1            & Dynamical evolution and population of MBHBs,
		study the birth and growth of MBHs, and
		the astrophysical environment of the host galaxy                         \\
		2            & Precisely estimate parameters of MBH,
		reveals the physical nature of BHs,
		probe dynamics of galactic nuclei, and
		the astrophysical environment at the galaxy center
		by EMRI                                                                  \\
		3            & Formation evolution and population of Galaxy binaries     \\
		4            & Study SOBHB formation, environment, population, and joint
		observation with LIGO                                                    \\
		5            & Test GR, study the properties of GW propagation           \\
		6            & GW cosmology, i.e. measurement of the Hubble constant
		and cosmology constant                                                   \\
		7            & PSD shape and upper limit of SGWB signals                 \\
		8            & Detect unmodeled signals                                  \\
		9            & New physics and cosmology beyond GR                       \\
		\bottomrule\end{tabularx}
\end{table}

GWs signals are expected to be many orders of magnitude below the noise level caused by laser frequency fluctuations \cite{amaro-seoane_laser_2017}.
A post-processing technique called TDI has been proposed to reduce laser noise to an acceptable level \cite{otto_time-delay_2015}.
The idea is to construct virtual equal-arm interference by combining the data streams in the appropriate way.
The second-generation TDI is applicable to rotating and flexing detectors with linearly varying arm lengths.
The actual Taiji orbits are affected by the gravitational field of the planets, which leads to non-periodic orbits.
In science analysis, it is necessary to extract scientific information from the various TDI data streams; in a similar manner to the procedure for the electromagnetic case, science analysis is carried out by deconvolution of the response function from the data stream in order to obtain the incident GW field observed by Taiji.

The main task of the data analysis is to quickly identify the signals and estimate their physical parameters precisely.
Unlike current ground-based detectors, Taiji will be sensitive to continuous sources, primarily from GBs.
It is expected that GBs will be the most numerous of the sources, the unresolved component of the galactic population produces an effectively stochastic "confusion noise".
Due to the non-Gaussianity of GB foreground noise, detecting other signals buried in the foreground noise is a challenging task.
What's more, analysis of the foreground itself is another challenge because of the enormous number of signals.
Because of the long duration and complex morphology of their signals, detecting and characterizing the physics of SOBBHs with LISA is a highly nontrivial challenge.
Data gaps will appear in the Taiji data stream due to maintenance, telescope realignment, and communication with Earth by a high-gain antenna.
A series of significant challenges arise when dealing with gapped data, namely the information loss, spectral leakage, and noise correlations \cite{blelly_sparse_2021}.
Another challenge for GW signal detection is the interference of instrumental glitches, which can be separated relatively natural from signals by ground observations with multiple interferometers.
In contrast, this coherent searching approach can not be directly applied in space-based detection, so novel glitch classification algorithms need to be developed \cite{robson_detecting_2019}.
All the technical challenges are summarized in Table \ref{tab:Technical challenge}.

\begin{table}[h!]
	\centering
	\caption{Technical challenges of Taiji mission}
	\label{tab:Technical challenge}
	\begin{tabularx}{0.5\textwidth}{@{} lYl @{}}
		\toprule
		\textbf{No.} & \textbf{Technical challenges}                                                \\
		\midrule
		1            & Foreground GB signals separation                                             \\
		2            & SOBBH signals separation                                                     \\
		3            & Others Signal with non-Gaussian GB noise                                     \\
		4            & Overlapping MBHB signals                                                     \\
		5            & Instrumental glitches                                                        \\
		6            & Data gap due to maintenance, telescope re-alignment, and communication issue \\
		7            & TDI technique to suppress laser frequency noise                              \\
		\bottomrule
	\end{tabularx}
\end{table}

\section{Method}
\label{sec:method}
When analyzing real detector data, it is necessary to model the signal and noise; therefore, for simplicity in the TDC, several idealized models are utilized during data simulation.
This section will introduce these models and their respective approximations.
Using the existing waveform model, we first obtain the source frame GW waveform during signal simulation.
Then, we determine the response of each link, which is dependent upon the orbit of the Taiji SCs.
In addition, TDI is used to obtain the signal in the X, Y, and Z channels.
The instrumental noise in those 3 channels is Gaussian noise simulated by their PSD.
In the final step, the signal is injected into the noise.

\subsection{Orbital motion}
We use Keplerian orbit to approximate the motion of the Taiji SC \cite{dhurandhar_fundamentals_2005,nayak_minimum_2006,wu_analytical_2019}. Given the distance between the SC i.e. the arm length $L$ and the distance from the SC to the sun $a$, the eccentricity of the orbit is defined by:
\begin{equation}
	\begin{aligned}
		e=\left(1+\frac{4}{\sqrt{3}} \alpha \cos \phi +\frac{4}{3} \alpha^{2}\right)^{1 / 2}-1,
	\end{aligned}
\end{equation}
where $\phi=\pi/3 + 5\alpha/8$ is the angle between the SC's orbital planes and the ecliptic plane. $\alpha = L/2a$ is a small parameter originally used in Ref. \cite{dhurandhar_fundamentals_2005}. The specific value of $\phi$ reduces the breathing effect of the arm length \cite{nayak_minimum_2006}.

The reference position of the 3 SC is given by:
\begin{align}
	\left\{
	\begin{aligned}
		x_{\mathrm{ref},k}(t) & = a \cos \iota (\cos \psi_k(t) - e),  \\
		y_{\mathrm{ref},k}(t) & = a \sqrt{1 - e^2} \sin \psi_k(t),    \\
		z_{\mathrm{ref},k}(t) & = -a \sin \iota (\cos \psi_k(t) - e).
	\end{aligned}
	\right.
\end{align}
where $\iota$ is the orbital inclination of SC1 relative to the ecliptic plane.
\begin{align}
	\tan \iota=\frac{2}{\sqrt{3}} \frac{\alpha \sin \phi}{\left(1+\frac{2}{\sqrt{3}} \alpha \cos \phi\right)},
\end{align}
where $\psi_k(t)$ is the eccentric anomaly of $\mathrm{SC}_k$, which is obtained by solving the Kepler equation to the desired order iteratively:
\begin{equation}
	\psi_k(t)-e \sin \psi_k(t)=m_k(t)=m_{0,k}+\Omega(t-t_0)=m_{0,1} - \frac{2 \pi (k-1)}{3} +\Omega(t-t_0), \quad \Omega = \frac{2\pi}{1\ \mathrm{year}}.
\end{equation}
Then, the position of 3 SC could be expressed in terms of the reference position and periastron $\lambda_k = \lambda_1 + 2 \pi (k-1) / 3$:
\begin{align}
	\left\{
	\begin{aligned}
		x_k(t) & = \cos \lambda_k x_{\mathrm{ref},k}(t) - \sin \lambda_k y_{\mathrm{ref},k}(t), \\
		y_k(t) & = \sin \lambda_k x_{\mathrm{ref},k}(t) + \cos \lambda_k y_{\mathrm{ref},k}(t), \\
		z_k(t) & = z_{\mathrm{ref},k}(t).
	\end{aligned}
	\right.
\end{align}

In summary, the orbits of 3 SCs are determined by 4 parameters: the semi-major axis $a$, the arm length $L$, $\rm{SC}_1$'s initial mean anomaly  $m_{0,1}$, and $\rm{SC}_1$'s initial periastron $\lambda_{1}$.

\subsection{Light travel time}
Following Ref. \cite{chauvineau_relativistic_2005}, we could use the Taylor expansion of the position of the emitter SC to calculate the light travel time analytically.
So the light travel time could be expressed as:
\begin{align}
	\mathrm{LTT} = \mathrm{LTT}^{(0)} + \mathrm{LTT}^{(1)} +\mathrm{LTT}^{(2)}+ \mathrm{LTT}^{(\mathrm{Sh})}
\end{align}
with
\begin{align}
	\mathrm{LTT}^{(0)}(t) & =\dfrac{1}{c}|\vec{x}_{\mathrm{recv}}(t)-\vec{x}_{\mathrm{send}}(t)|                                                                                                                               \\
	\mathrm{LTT}^{(1)}(t) & =\dfrac{\vec{v}_{\mathrm{recv}}(t)}{c^2}\cdot (\vec{x}_{\mathrm{recv}}(t)-\vec{x}_{\mathrm{send}}(t))                                                                                              \\
	\mathrm{LTT}^{(2)}(t) & =\dfrac{|\vec{x}_{\mathrm{recv}}(t)-\vec{x}_{\mathrm{send}}(t)|}{c^3} \bigg[ \vec{v}_{\mathrm{recv}}^2(t) - \vec{a}_{\mathrm{recv}}(t)\cdot(\vec{x}_{\mathrm{recv}}(t)-\vec{x}_{\mathrm{send}}(t)) \\
	                      & \qquad + \left(\frac{\vec{v}_{\mathrm{recv}}(t)\cdot(\vec{x}_{\mathrm{recv}}(t)-\vec{x}_{\mathrm{send}}(t))}{|\vec{x}_{\mathrm{recv}}(t)-\vec{x}_{\mathrm{send}}(t)|}\right)^2 \bigg],
\end{align}
where the subscripts 'send' and 'recv' represent the emitter SC and the receiver SC respectively.
Finally, we consider the Shapiro delay caused by the Sun's gravitational field, which is a second-order effect \cite{hees_relativistic_2014}:
\begin{equation}
	\mathrm{LTT}^{(\mathrm{Sh)}} = \frac{R_S}{c}\cdot\ln{ \frac{|\vec{x}_{\mathrm{send}}(t)| + |\vec{x}_{\mathrm{recv}}(t)| + |\vec{x}_{\mathrm{recv}}(t)-\vec{x}_{\mathrm{send}}(t)| }{|\vec{x}_{\mathrm{send}}(t)| + |\vec{x}_{\mathrm{recv}}(t)| - |\vec{x}_{\mathrm{recv}}(t)-\vec{x}_{\mathrm{send}}(t)|}},
\end{equation}
where $R_S$ is the Schwarzschild radius of the Sun.

\subsection{Detector response}
\label{subsec: resp}
For the data generation, we use a time domain detector response and the GPU accelerated code provided by \cite{katz_assessing_2022}, which enables us to generate a large amount of data directly in the time domain.
First, we transform the GW waveform from the source frame to the solar-system barycenter (SSB) frame by
\begin{equation}
	\left[\begin{array}{c}
			h_{+}^\mathrm{SSB} \\
			h_{\times}^\mathrm{SSB}
		\end{array}\right]=\left[\begin{array}{cc}
			\cos 2 \psi & -\sin 2 \psi \\
			\sin 2 \psi & \cos 2 \psi
		\end{array}\right]\left[\begin{array}{l}
			h_{+}^\mathrm{src} \\
			h_{\times}^\mathrm{src}
		\end{array}\right] \text {. }
\end{equation}
Then, the length variation of one arm caused by the GW signal is:
\begin{align}
	H(t) = h_{+}^{\mathrm{SSB}} (t) \xi_{+} (\hat{\theta},\hat{\phi},\vec{n}(t)) + h_{\times}^{\mathrm{SSB}} (t) \xi_{\times} (\hat{\theta},\hat{\phi},\vec{n}(t)),
\end{align}
where the antenna pattern $\xi_{+}$ and $\xi_{\times}$ are given by:
\begin{equation}
	\left\{
	\begin{aligned}
		\xi_{+}(\vec{u}, \vec{v}, \vec{n})      & =  {\left(\vec{u} \cdot \vec{n} \right)}^2 - {\left(\vec{v}\cdot \vec{n} \right)}^2, \\
		\xi_{\times}(\vec{u}, \vec{v}, \vec{n}) & =  2 \left(\vec{u} \cdot \vec{n} \right) \left(\vec{v} \cdot \vec{n} \right),
	\end{aligned}
	\right.
\end{equation}
$\hat{n}(t)$ is the unit vector of the arm, $\hat{\theta}, \hat{\phi}$ are polar and azimuthal angle in the SSB frame, based on the orthonormal basis vectors $(\hat{e}_r,\hat{e}_{\theta},\hat{e}_{\phi})$.
$\vec{u}, \vec{v}$ are polarization vectors defined by $\vec{u}=-\hat{e}_{\phi}$ and $\vec{v}=-\hat{e}_{\theta}$. The propagation vector is $\vec{k}=-\hat{e}_{r}$.
Light travel time along the arm affected by the GW is :
\begin{align}
	\label{eq:t_gw}
	t_{\mathrm{recv}} \simeq t_{\mathrm{send}} + \mathrm{LTT} - \frac{1}{2c} \int_0^L H(\vec{x}(\lambda),t(\lambda))\, d\lambda,
\end{align}
where we use first order approximation as $t(\lambda)=t_{\mathrm{send}}+\lambda/c, \vec{x}(\lambda)=\vec{x}_{\mathrm{send}}(t_{\mathrm{send}})+\lambda \vec{n}(t_{\mathrm{send}})$, $\vec{x}_{\mathrm{send}}$ is the position of the
emitter SC.
The parameter $\lambda$ used as the integration variable describes the photon path.
With those approximations, $H(\vec{x}(\lambda),t(\lambda))$ is given by:
\begin{equation}
	\label{eq:H_x_t}
	H(\vec{x}(\lambda),t(\lambda))=H\left(t(\lambda) - \frac{\vec{k}\cdot\vec{x}(\lambda)}{c}\right)=H\left(t_{\mathrm{send}} - \frac{\vec{k}\cdot\vec{x}_{\mathrm{send}} (t_{\mathrm{send}})}{c}
	+ \lambda \frac{1-\vec{k}\cdot\vec{n}(t_{\mathrm{send}})}{c}\right)
\end{equation}

Combining equation. \eqref{eq:t_gw} and \eqref{eq:H_x_t}  and differentiating the resulting expression with respect to $t_{\mathrm{send}}$ yields the relative frequency shift,
\begin{align}
	y (t_{\mathrm{send}}) = \frac{1}{2(1 - \vec{k}\cdot\vec{n}(t_{\mathrm{send}}))}\left[ H \left( t_{\mathrm{send}} - \frac{\vec{k}\cdot\vec{x}_{\mathrm{send}}(t_{\mathrm{send}})}{c} \right) -  H \left( t_{\mathrm{send}} - \frac{\vec{k}\cdot\vec{x}_{\mathrm{recv}}(t_{\mathrm{recv}})}{c} + \mathrm{LTT}\right) \right].
\end{align}
We assume that $\vec{n}(t_{\mathrm{send}})\simeq \vec{n}(t_{\mathrm{recv}})$ since the variation of arm direction due to the SC's motion during the light travel time is less than $10^{-3}\mathrm{rad}$ in the case of Taiji \cite{rijnveld_picometer_2017} and $t_{\mathrm{recv}}\simeq t_{\mathrm{send}} + \mathrm{LTT}$. Finally, the GW strain of one arm is:
\begin{align}
	y (t_{\mathrm{recv}}) = \frac{1}{2(1 - \vec{k}\cdot\vec{n}(t_{\mathrm{recv}}))}\left[ H \left( t_{\mathrm{recv}} - \frac{\vec{k}\cdot\vec{x}_{\mathrm{send}}(t_{\mathrm{recv}})}{c} - \mathrm{LTT}(t_{\mathrm{recv}}) \right) -  H \left( t_{\mathrm{recv}} - \frac{\vec{k}\cdot\vec{x}_{\mathrm{recv}}(t_{\mathrm{recv}})}{c} \right) \right].
\end{align}

\subsection{TDI Combination}
\label{subsec: tdi}
Due to the unequal arm length of the Taiji SCs, the laser frequency noise could not be canceled naturally, which will be orders of magnitude above the GW signals.
So the TDI technique was developed to cancel the noise by shifting the data from different SCs by subtle time delays and combining them together.
The TDI technique can be divided into several generations:
\begin{itemize}
	\item TDI 1.0 is used in the static case, which does not consider the rotation and breathing effect. In other words, the arm lengths are constants in time, i.e. the light travel time satisfies $\mathrm{LTT}_i=\mathrm{LTT}_{i^{\prime}}=const$.
	\item TDI 1.5 is used in the rigid rotating case, which does not consider the breathing effect.
	      The arm lengths are still constants in time, i.e. $\mathrm{LTT}_i=const., \mathrm{LTT}_{i^{\prime}}=const.$ but $\mathrm{LTT}_i \neq \mathrm{LTT}_{i^{\prime}}$ here.
	\item TDI 2.0 is used in the flexing case, the arm lengths are functions of time, i.e. $\mathrm{LTT}_i=\mathrm{LTT}_i(t), \mathrm{LTT}_{i^{\prime}}=\mathrm{LTT}_{i^{\prime}}(t)$, and $\mathrm{LTT}_i(t)\neq \mathrm{LTT}_{i^{\prime}}(t)$.
\end{itemize}
Here we adopt the notation from Ref. \cite{otto_time-delay_2015}, the number in the subscript represents different links, where $i\in\{1,2,3\}$ denotes the clockwise link opposite to $\rm{SC}_i$ and $i^\prime$ denotes the counter-clockwise link.
The first generation unequal arm Michelson combination is:
\begin{equation}
	\label{eq:TDI1.0}
	\begin{aligned}
		X_1(t)=\left(y_{2^{\prime}: 322^{\prime}}+y_{1: 22^{\prime}}+y_{3: 2^{\prime}}+y_{1^{\prime}}\right)-\left(y_{3: 2^{\prime} 3^{\prime} 3}+y_{1^{\prime}: 3^{\prime} 3}+y_{2^{\prime}: 3}+y_1\right). \\
	\end{aligned}
\end{equation}

The second-generation Michelson combination is
\begin{equation}
	\label{eq:TDI2.0}
	\begin{aligned}
		 & X_2(t)=y_{1^{\prime}}+y_{3 , 2^{\prime}}+y_{1 , 22^{\prime}}+y_{2^{\prime} , 322^{\prime}}+y_{1 , 3^{\prime} 322^{\prime}}+y_{2^{\prime} , 33^{\prime} 322^{\prime}}                                    \\
		 & \quad +y_{1^{\prime} , 3^{\prime} 33^{\prime} 322^{\prime}}+y_{3 , 2^{\prime} 3^{\prime} 33^{\prime} 322^{\prime}}-y_1-y_{2^{\prime} , 3}-y_{1^{\prime} , 3^{\prime} 3}-y_{3 ; 2^{\prime} 3^{\prime} 3} \\
		 & \quad-y_{1^{\prime} , 22^{\prime} 3^{\prime} 3}-y_{3 , 2^{\prime} 22^{\prime} 3^{\prime} 3}-y_{1 ; 22^{\prime} 22^{\prime} 3^{\prime} 3}-y_{2^{\prime} , 322^{\prime} 22^{\prime} 3^{\prime} 3}.
	\end{aligned}
\end{equation}
The variables $Y$ and $Z$ could be obtained by cyclic permutation of the indices. The complete derivation and detailed explanation of these equations can be found in Ref. \cite{otto_time-delay_2015}.
The colon denotes time delay when the arm length is time-independent,
\begin{equation}
	f(t)_{: j}:=f\left(t-\frac{L_j}{c}\right), \quad f(t)_{: j k}=f\left(t-\frac{L_k}{c}-\frac{L_j}{c}\right) =f(t)_{: kj},
\end{equation}
which is commute for $j\neq k$. Next, a comma denotes time delays with time-dependent arm lengths, here we compute chained delays as simple sums of delays rather than nested delays,
\begin{equation}
	f(t)_{, j k}=f\left(t-\frac{L_k(t)}{c}-\frac{L_j(t)}{c}\right).
\end{equation}
This approximation is sufficient for computing the GW response function \cite{katz_assessing_2022}.

After performing the TDI combination, we obtain the X, Y, and Z channel gravitational wave strain data. Simulating the instrumental noise is another important aspect of the data simulation process, distinct from the strain data. In the next subsection, we will delve into the detailed process of simulating instrumental noise.

\subsection{Instrumental noise}
Despite the complex noise sources that will be involved in the Taiji detection data in practice, in our simulation, we use a noise model consisting of just two components for simplicity.
The first one is the high-frequency component, which represents the overall noises that come from the optical metrology system, and the second is the low-frequency component, which represents the test mass acceleration noise.
The power spectral density (PSD) is given by:
\begin{equation}
	\begin{aligned}
		P_{\mathrm{oms}}(f) & = 64\times 10^{-24} \frac{1}{\mathrm{Hz}}\left[1+\left(\frac{2 \mathrm{mHz}}{f}\right)^4\right]\left(\frac{2 \pi f}{c}\right)^2,                                                             \\
		P_{\mathrm{acc}}(f) & = 9\times 10^{-30}\frac{1}{ \mathrm{Hz}}\left[ 1+\left(\frac{0.4 \mathrm{mHz}}{f}\right)^2\right]\left[1+\left(\frac{f}{8 \mathrm{mHz}} \right)^4 \right] \left(\frac{1}{2 \pi fc}\right)^2.
	\end{aligned}
\end{equation}
Following \cite{babak_lisa_2021}, the analytic model of the one-sided noise PSD is derived from the noise components in each interferometric measurement.
We will explicitly show PSD for a Michelson-type TDI generator.
Given the data combination \eqref{eq:TDI1.0} and \eqref{eq:TDI2.0} with the assumption that the noise of each link has the same PSD, the noise PSD of the first and the second generation TDI X channel is:
\begin{equation}
	\mathrm{PSD}_{X_{1}}=16 \sin ^2(\omega L)\left(P_{\mathrm{oms}}+(3+\cos (2 \omega L)) P_{\mathrm{acc}}\right),
\end{equation}
\begin{equation}
	\mathrm{PSD}_{X_{2}}=64 \sin ^2(\omega L) \sin ^2(2 \omega L)\left(P_{\mathrm{oms}}+(3+\cos (2 \omega L)) P_{\mathrm{acc}}\right),
\end{equation}
where $\omega = 2\pi f/c$. In this paper, we use TDI 2.0 PSD for noise generation.

\section{Dataset}
\label{sec:data}
\subsection{Overall setting}
The TDC dataset was designed based on the scientific objectives of Taiji.
For most of the sources of Taiji, we do not have well-established data analysis algorithms because they are not detected by ground-based detectors.
So at the beginning of the data challenge, we should focus on each source separately for simplicity.
Each of the datasets from TDC-1 to TDC-6 contains signals from a single source.
Multiple sources are included in the last dataset TDC-7, including MBHBs, VGBs, and GB foregrounds.
Those datasets cover all the sources for the scientific objectives.
Similar to LDC, the sampling rate here in TDC is 0.1 Hz for all datasets, while this setting will be adjustable with Taiji's configuration accordingly.
First, we generate the signal, then compute the TDI 2.0 response, and add it to the simulated noise.
For the time domain signals, we use the time domain TDI response as mentioned in \ref{subsec: resp} and \ref{subsec: tdi}. For the stochastic GW signals, we generate them directly by their PSD, which considers TDI response.
Next, we will introduce the configuration and parameters used to generate the datasets.

\begin{figure}[htbp]
	\centering
	\begin{subfigure}[t]{0.3\textwidth}{
			\includegraphics[width=\textwidth,valign=t]{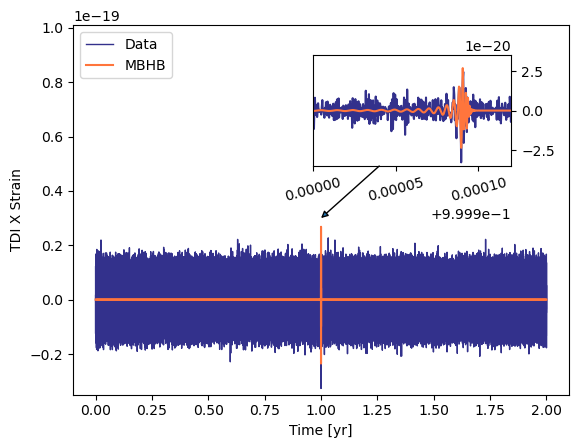}
			\caption{\label{fig:tdc1td}}}
	\end{subfigure}
	\begin{subfigure}[t]{0.3\textwidth}{
			\includegraphics[width=\textwidth,valign=t]{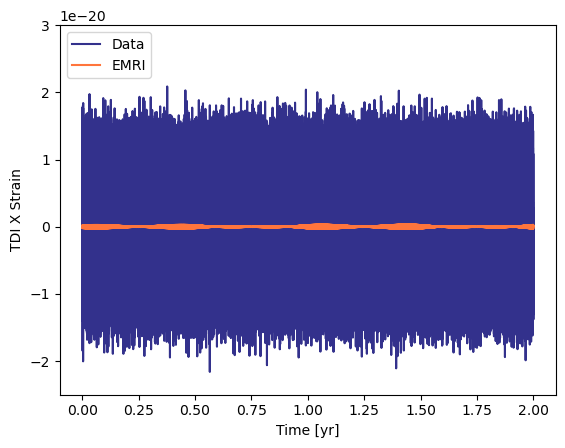}
			\caption{\label{fig:tdc21td}}}
	\end{subfigure}
	\begin{subfigure}[t]{0.3\textwidth}{
			\includegraphics[width=\textwidth,valign=t]{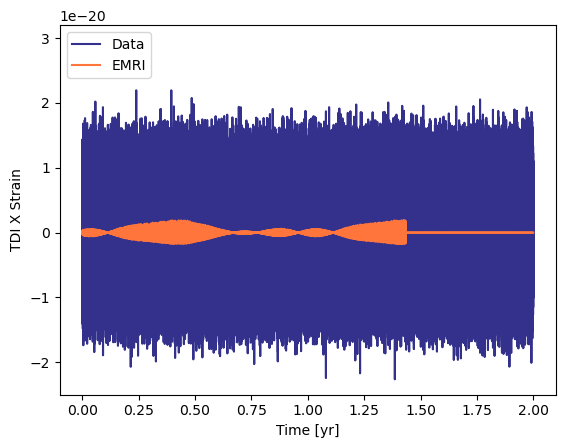}
			\caption{\label{fig:tdc22td}}}
	\end{subfigure}
	\begin{subfigure}[t]{0.3\textwidth}{
			\includegraphics[width=\textwidth,valign=t]{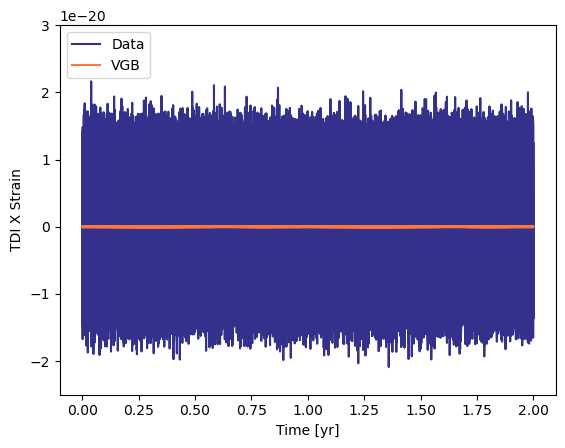}
			\caption{\label{fig:tdc3td}}}
	\end{subfigure}
	\begin{subfigure}[t]{0.3\textwidth}{
			\includegraphics[width=\textwidth,valign=t]{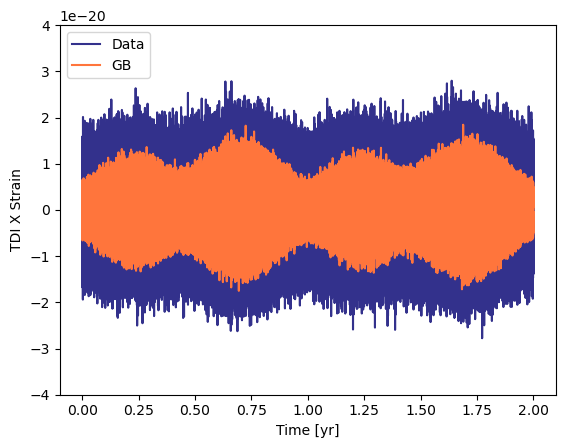}
			\caption{\label{fig:tdc4td}}}
	\end{subfigure}
	\begin{subfigure}[t]{0.3\textwidth}{
			\includegraphics[width=\textwidth,valign=t]{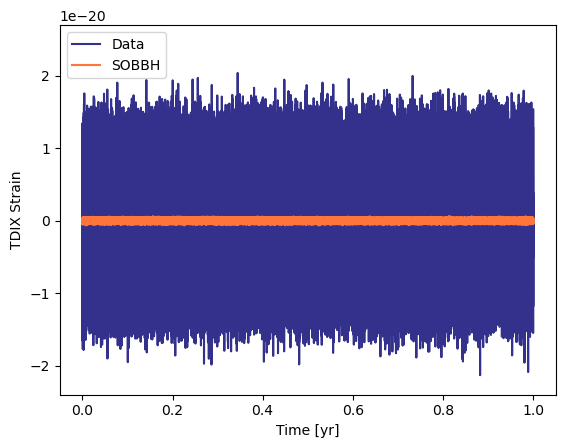}
			\caption{\label{fig:tdc5td}}}
	\end{subfigure}
	\begin{subfigure}[t]{0.3\textwidth}{
			\includegraphics[width=\textwidth,valign=t]{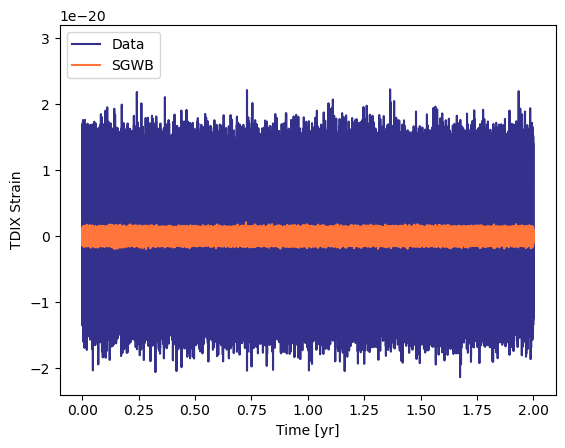}
			\caption{\label{fig:tdc61td}}}
	\end{subfigure}
	\begin{subfigure}[t]{0.3\textwidth}{
			\includegraphics[width=\textwidth,valign=t]{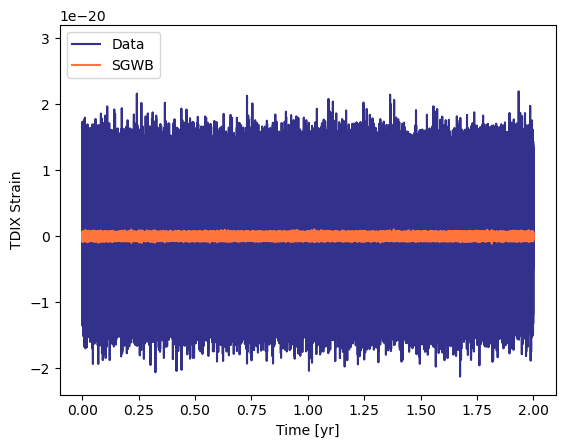}
			\caption{\label{fig:tdc62td}}}
	\end{subfigure}
	\begin{subfigure}[t]{0.3\textwidth}{
			\includegraphics[width=\textwidth,valign=t]{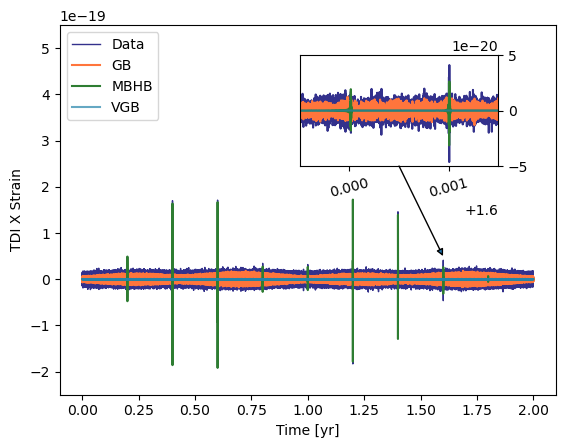}
			\caption{\label{fig:tdc7td}}}
	\end{subfigure}
	\caption{Time-domain data of all the TDC datasets. The strain of TDI X channel data and signals are presented. (a), TDC-1, the MBHB signal is zoomed in because of its duration. (b) and (c), TDC-2-1 and TDC-2-2. (d), TDC-3. (e), TDC-4. (f), TDC-5. (g) and (h), TDC-6-1 and TDC-6-2. (i), TDC-7, signals No.8 and No.9 are zoomed in due to the closeness of their coalescence time.}
	\label{fig:td}
\end{figure}

\begin{figure}[htbp]
	\centering
	\begin{subfigure}[t]{0.3\textwidth}{
			\includegraphics[width=\textwidth,valign=t]{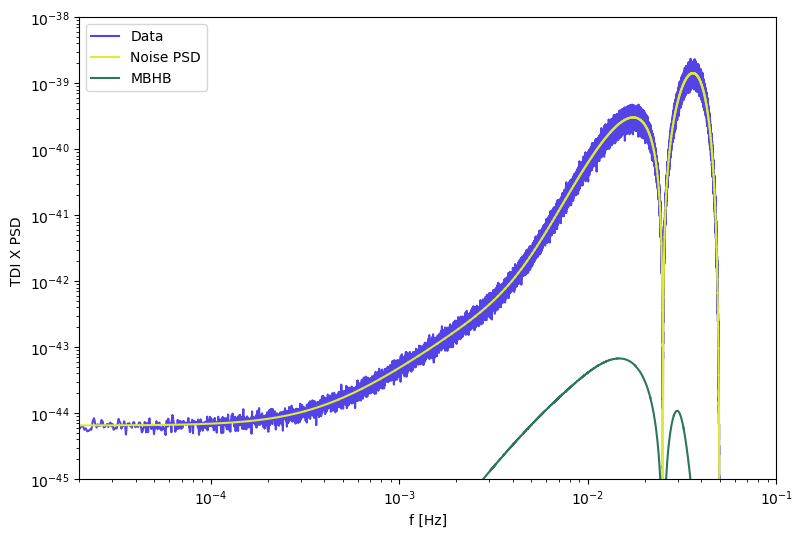}
			\caption{\label{fig:tdc1fd}}}
	\end{subfigure}
	\begin{subfigure}[t]{0.3\textwidth}{
			\includegraphics[width=\textwidth,valign=t]{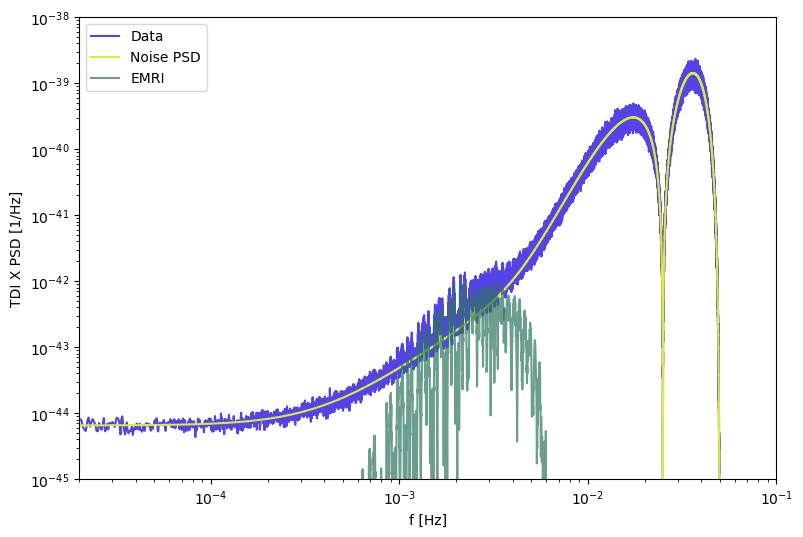}
			\caption{\label{fig:tdc21fd}}}
	\end{subfigure}
	\begin{subfigure}[t]{0.3\textwidth}{
			\includegraphics[width=\textwidth,valign=t]{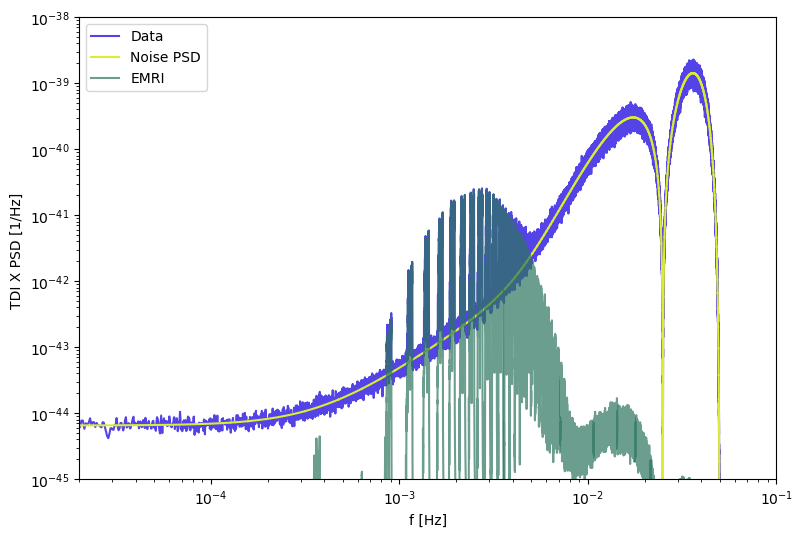}
			\caption{\label{fig:tdc22fd}}}
	\end{subfigure}
	\begin{subfigure}[t]{0.3\textwidth}{
			\includegraphics[width=\textwidth,valign=t]{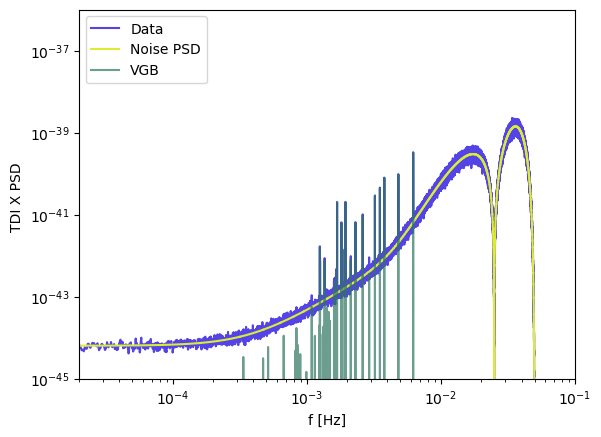}
			\caption{\label{fig:tdc3fd}}}
	\end{subfigure}
	\begin{subfigure}[t]{0.3\textwidth}{
			\includegraphics[width=\textwidth,valign=t]{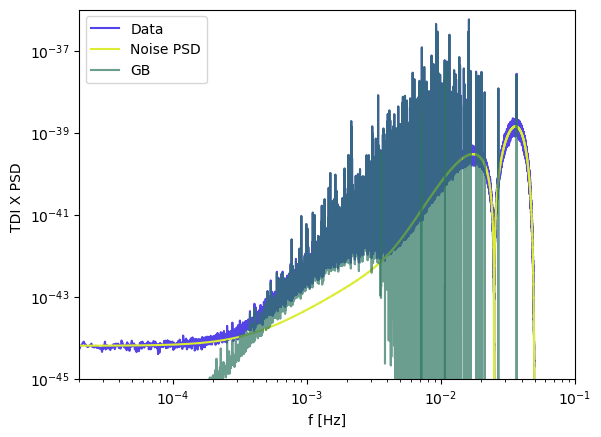}
			\caption{\label{fig:tdc4fd}}}
	\end{subfigure}
	\begin{subfigure}[t]{0.3\textwidth}{
			\includegraphics[width=\textwidth,valign=t]{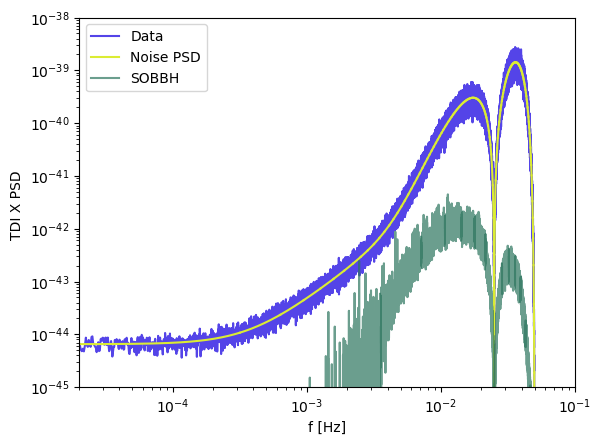}
			\caption{\label{fig:tdc5fd}}}
	\end{subfigure}
	\begin{subfigure}[t]{0.3\textwidth}{
			\includegraphics[width=\textwidth,valign=t]{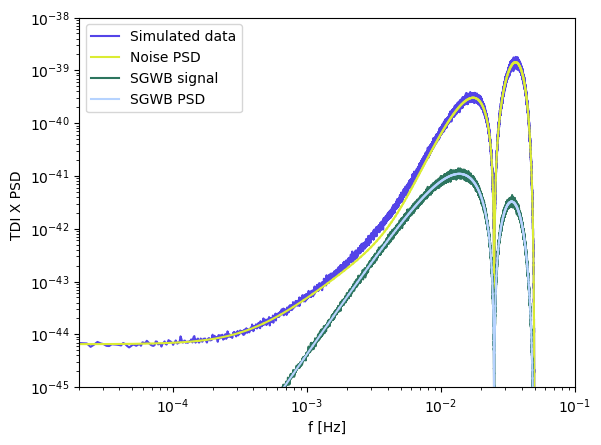}
			\caption{\label{fig:tdc61fd}}}
	\end{subfigure}
	\begin{subfigure}[t]{0.3\textwidth}{
			\includegraphics[width=\textwidth,valign=t]{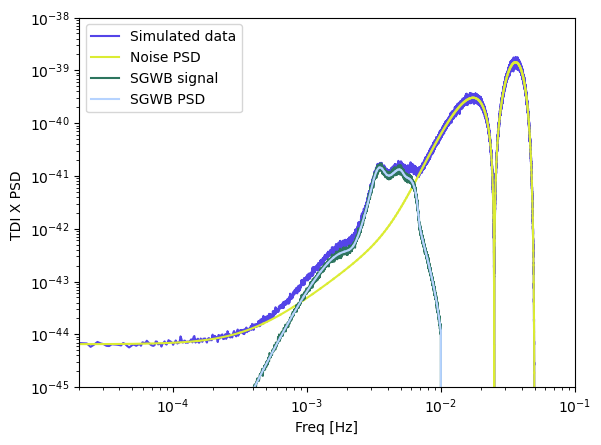}
			\caption{\label{fig:tdc62fd}}}
	\end{subfigure}
	\begin{subfigure}[t]{0.3\textwidth}{
			\includegraphics[width=\textwidth,valign=t]{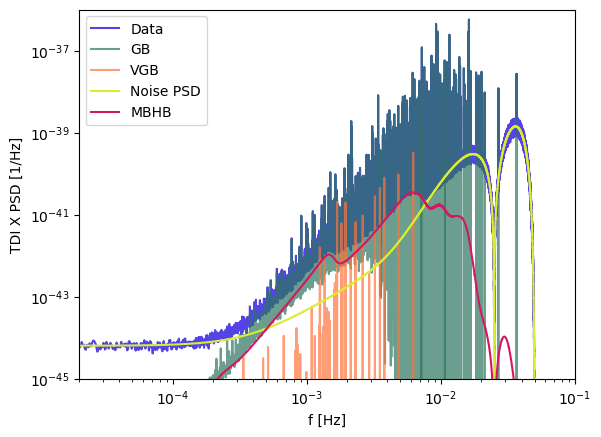}
			\caption{\label{fig:tdc7fd}}}
	\end{subfigure}
	\caption{Frequency domain data of all the TDC datasets. The PSD of TDI X channel data and signals are compared with the noise PSD. (a), TDC-1. (b) and (c), TDC-2-1 and TDC-2-2. (d), TDC-3. (e), TDC-4. (f), TDC-5. (g) and (h), TDC-6-1 and TDC-6-2. (i), TDC-7.}
	\label{fig:fd}
\end{figure}

\subsection{Massive black hole binary}
MBHBs are primary sources of Taiji, which is sensitive to MBHBs with total mass in the range of $[10^5,10^8]M_\odot$ \cite{amaro-seoane_laser_2017}.
It is expected to detect $1\sim 20$ MBHBs per year \cite{katz_probing_2019}.
We will be able to gain valuable insight into how MBHs are formed by measuring their spins and masses.
It should be possible for us to test the hypothesis that modern galaxies are the result of the merger of smaller "seed" galaxies in the early universe, as we will be able to detect the merger of MBHB out to a high redshift.
We use SEOBNRv4\_{opt} \cite{bohe_improved_2017} here to generate the MBHB signal and shift it by the coalescence time $t_c$.
Only the $l=m=2$ mode is included in the data. Both TDC-1 and TDC-7 contain MBHB signals.
For TDC-1, there is only one MBHB signal in this dataset, the time domain and the frequency domain behavior of the TDC-1 data are shown in Fig. \ref{fig:tdc1td} and Fig. \ref{fig:tdc1fd} respectively.
In the time domain, we could see that the merger portion of the signal is not buried in the noise, which has a short duration of $\sim$ hour.
Then, in the frequency domain, the signal is below the noise PSD. The parameter of MBHB signals injected in those two datasets are shown in Table \ref{tab:mbhb_par}, where $M_T$ is the total mass of two black holes, $q$ is the mass ratio with $q<1$, $s_1, s_2$ is the spin parameters, $\iota$ is the inclination angle, $\psi$ is the polarization angle, $\lambda$, and $\beta$ is ecliptic longitude and ecliptic latitude characterize the sky location of the source, $t_c$ is the coalescence time, and $\phi_c$ is coalescence phase.
TDC-1 only contains signal No. 5 and all 10 signals are included in TDC-7.
The time domain and the frequency domain behavior of the TDC-7 data are shown in Fig. \ref{fig:tdc7td} and Fig. \ref{fig:tdc7fd} respectively. Signals No. 8 and No. 9 are close together, presenting additional challenges to data processing algorithms that are not encountered by ground-based detectors.

\begin{table}[h!]
	\centering
	\caption{Parameters of MBHB signals used in TDC-1 and TDC-7}
	\label{tab:mbhb_par}
	\begin{tabular}{@{} lccccccccccc @{}}
		\toprule
		No. & $M_T(M_\odot)$   & $q$  & $s_1$ & $s_2$ & $D_L$(Gpc) & $\iota$    & $\psi$   & $\lambda$ & $\beta$    & $t_c(yr)$ & $\phi_c$ \\
		\midrule
		1   & $1.2\times 10^6$ & 0.3  & 0.1   & 0.2   & 53.39      & $2\pi/3$   & $\pi/3$  & $\pi/10$  & $\pi/8$    & $0.2$     & 1        \\
		2   & $1.8\times 10^6$ & 0.8  & 0.14  & 0.12  & 16.02      & $\pi$      & $\pi/3$  & $\pi/15$  & $5\pi/4$   & $0.4$     & 0.05     \\
		3   & $2.4\times 10^6$ & 0.4  & 0.18  & 0.08  & 21.36      & $5\pi/6$   & $\pi/3$  & $\pi/20$  & $7\pi/4$   & $0.6$     & 0.1      \\
		4   & $1.5\times 10^6$ & 0.7  & 0.14  & 0.8   & 53.39      & $\pi/3$    & $\pi/3$  & $\pi/25$  & $3\pi/4$   & $0.8$     & 0.15     \\
		5   & $6\times 10^5$   & 0.5  & 0.2   & 0.4   & 53.39      & $\pi/3$    & $\pi/3$  & $\pi/5$   & $\pi/4$    & $1$       & 0.5      \\
		6   & $1.2\times 10^6$ & 0.6  & 0.1   & 0.6   & 106.78     & $\pi/6$    & $2\pi/3$ & $\pi/30$  & $5\pi/4$   & $1.2$     & 0.35     \\
		7   & $2.4\times 10^6$ & 0.1  & 0.22  & 0.48  & 16.02      & $4\pi/15$  & $5\pi/3$ & $\pi/35$  & $\pi/8$    & $1.4$     & 0.45     \\
		8   & $1.2\times 10^6$ & 0.05 & 0.16  & 0.28  & 26.69      & $13\pi/30$ & $\pi/3$  & $2\pi/5$  & $\pi/12$   & $1.6$     & 1        \\
		9   & $1.8\times 10^6$ & 0.5  & 0.02  & 0.08  & 32.03      & $14\pi/30$ & $\pi/3$  & $3\pi/5$  & $11\pi/40$ & $1.601$   & 0.4      \\
		10  & $1.2\times 10^7$ & 1    & 0.06  & 0.08  & 10.68      & $\pi/3$    & $\pi/3$  & $6\pi/5$  & $\pi/4$    & $1.8$     & 0.5      \\
		\bottomrule
	\end{tabular}
\end{table}

\subsection{Extreme-mass-ratio inspiral}
EMRIs are binary black hole systems consisting of a stellar-mass compact object (CO) and a massive black hole with mass $M \sim 10^4 M_\odot - 10^7 M_\odot$. Taiji will observe few tens to hundreds EMRI events over 2 years with signal-to-noise ratio (SNR) above 20 \cite{gair_prospects_2017}. Because of the extremely large mass ratio, the EMRI evolves slowly, inspiraling about $10^4\sim 10^5$ cycles in Taiji's frequency band \cite{babak_science_2017}. Since EMRI signals contain rich information about the geometry surrounding the central black holes, they are one of the primary fundamental physics goals of the Taiji mission. It has been suggested that detections of EMRIs could provide a highly accurate observational test of the "Kerr-ness" of the central MBH \cite{barack_using_2007}.What's more, EMRIs sample the stellar population near the central black holes.By detecting EMRI signals, we can probe the physical properties of the central MBH and test GR  \cite{amaro-seoane_laser_2017}.
Several waveform templates are developed for such a promising source of Taiji \begin {enumerate*} [label=\itshape\arabic*\upshape)]\item Teukolsky-based method, \cite{glampedakis_extreme_2005,han_constructing_2011}\item Numerical Kludge (NK)\cite{babak_kludge_2007},    \item Analytic Kludge (AK)\cite{barack_lisa_2004},     \item Augmented Analytic Kludge (AAK) \cite{chua_improved_2015,chua_fast_2017, katz_fastemriwaveforms_2021}, and    \item XSPEG \cite{xin_gravitational_2019,zhang_analytical_2021}.\end {enumerate*}  The first 2 of them are computationally expensive, and the AK waveform doesn't suit Kerr space-time, so we utilize the GPU-accelerated AAK waveform \cite{katz_fastemriwaveforms_2021} for the TDC-2-1 dataset generation and the XSPEG waveform for the TDC-2-2 dataset generation.  There is only one EMRI signal in each dataset. Because of the KRZ metric used in XSPEG waveform construction, TDC-2-2 could be used to test GR, which is a key scientific objective of the Taiji mission.  The time domain and the frequency domain behavior of the data are shown in Fig. \ref{fig:tdc21td} and \ref{fig:tdc22fd}.Unlike MBHBs, the amplitude of a typical EMRI is an order of magnitude below the instrumental noise. Detection is only possible if enough SNR is accumulated over a number of wave cycles. The parameters of EMRI waveform injected in TDC-2-1 is $(M, \mu, a_0, e_0, p_0, \iota_0, \theta_S, \phi_S, \theta_K, \phi_K, \Phi_{\varphi,0}, \Phi_{\theta,0}, \Phi_{r,0}, D_L)=(10^6 M_\odot, 30M_\odot, 0.6, 0.6, 15, 0.7, 10^{-6}, 10^{-6}, 10^{-6}, 10^{-6}, 0, 0, 0, 100\mathrm{Mpc})$, $M$ and $a$ are the mass and spin parameter of the MBH, $p$ is the semi-latus rectum, $e$ is eccentricity, $\iota$ is the orbit’s inclination angle from the equatorial plane, $\theta_S$, and $\phi_S$ are the polar and azimuthal sky location angles. $\theta_K$ and $\phi_K$ are the azimuthal and polar angles describing the orientation of the spin angular momentum vector of the MBH. $\Phi_{\varphi,0}, \Phi_{\theta,0}, \Phi_{r,0}$, represent the phase of azimuthal, polar, and radial modes respectively, $D_L$ is the luminosity distance. The parameters of the EMRI waveform injected in TDC-2-2 is $(M, \mu, a_0, e_0, p_0, \iota_0, \lambda, \beta, D_L)=(10^6 M_\odot, 10 M_\odot, 0.9, 0.6, 15, \pi/4, 0, \pi/4, 8.3\mathrm{Mpc})$, and the KRZ metric deformation parameter is $(\delta_1, \delta_2, \delta_3) = (0.1, 0.1, 0.1)$. Compared to the existing LDC1-2 dataset for EMRI, an updated AAK waveform template is used in TDC-2-1, and a waveform based on the KRZ metric is used in TDC-2-2.

\subsection{Galactic binary}
We use the simple GB model from Ref. \cite{katz_assessing_2022}, which is expressed in the source frame as:\begin{equation}
	\begin{aligned}
		\label{bwd_wf}
		h_{+}^{\mathrm{src}}(t)    & =\mathcal{A}(1+\cos^2\iota)\cos\Phi(t),                             \\
		h_\times^{\mathrm{src}}(t) & = 2\mathcal{A}\sin\iota\sin\Phi(t),                                 \\
		\Phi(t)                    & =\phi_0+2\pi f_0 t +\pi \dot{f_0} t^2 + \frac{\pi}{3}\Ddot{f_0}t^3, \\
		\Ddot{f_0}                 & =\frac{11}{3}\frac{\Dot{f_0^2}}{f_0}.
	\end{aligned}
\end{equation}
where $\mathcal{A}$ is the overall amplitude, $\phi_0$ is the initial phase at the start of the observation, $\iota$ is the inclination of the BWD orbit to the line of sight from the origin of the SSB frame. The intrinsic parameter is the frequency of the signal $f$ and its derivative $\dot{f}$. The comparison to LDC's GB waveform $\Ddot{f_0}$ is considered here.There are enormous GBs populated in the Milky Way \cite{amaro-seoane_laser_2017} which form foreground noise, and disentangling those signals is a challenging task \cite{zhang_resolving_2021}. With the help of GWs, GB foreground observations provide a key method for measuring the properties of a fraction of the galactic stellar population that is routinely assumed but has never been directly observed. What's more, there are some bright GBs called LISA verified binaries \cite{10.1093/mnras/sty1545}, the latest catalog is available online\footnote{\url{https://gitlab.in2p3.fr/LISA/lisa-verification-binaries}}. Following the above catalog, we generate the TDC-3 dataset, which contains 43 VGBs. The foreground GB noise is in the TDC-4 dataset, which consists of $\sim 3e7$ GB signals. The catalog of the TDC-4 dataset comes from the population study of GBs \cite{nelemans_gravitational_2001,nelemans_reconstructing_2005}.Both the signal and the noise PSD of these two datasets are shown in Fig. \ref{fig:tdc3fd} and \ref{fig:tdc4fd}, associated with the corresponding time domain data shown in \ref{fig:tdc3td} and \ref{fig:tdc4td}. Compared to LDC1's dataset, we use TDI 2.0 here. That GB noise is also included in TDC-7. The Taiji mission faces a data analysis challenge because of the presence of non-Gaussian and non-stationary astrophysical foregrounds. Due to the presence of foreground signals, GW source separation procedures will be complicated. A resolvable signal's source parameter estimation will also be affected by the level of confusion noise. Therefore, it is critical to study, understand, and predict the overall shape and amplitude of the potential foreground components of the Taiji data.

\subsection{Stellar-origin binary black holes}
Space-based detectors observe the early inspiral of SOBBH systems, whose merger will be detected by ground-based detectors.
Those systems evolve slowly in the LISA and Taiji frequency bands, which enable high-precision parameter estimation \cite{bambi_space-based_2021}.
The joint observation of SOBBH by Taiji and ground-based detectors could be used to probe low-frequency modifications due to deviations from GR or to environmental effects, to facilitate electromagnetic joint observations \cite{toubiana_parameter_2020,sesana_prospects_2016}.
By estimating coalescence time and sky location, we could forecast the signals for ground-based detectors.
We generate the TDC-5 dataset, which contains 21721 SOBBH signals, whose catalog comes from \cite{sesana_prospects_2016}.
The time domain and the frequency domain behavior of the data are shown in Fig. \ref{fig:tdc5td} and \ref{fig:tdc5fd}.
The SOBBH signals are completely below the noise in the frequency domain, which are background signals.
We generate the waveform using the IMRPhenomD template, then transform it to the time domain and apply the TDI response.
The main difference between the TDC-5 and LDC1-5 datasets is that we use the TDI 2.0 response in the time domain rather than the TDI 1.0 response in the frequency domain.

\subsection{Stochastic gravitational wave background}
One of the main scientific objectives of Taiji is to probe GWs from the early universe and reveal physics of GW sources \cite{bartolo_probing_2022}. Fundamental physics and astronomy would be greatly enhanced by the detection of a cosmological background from the early universe. The only way to obtain direct information about earlier epochs than decoupling may be through GWs and possibly neutrinos. There are many possible astrophysical and cosmological sources that contribute to the stochastic background, and up to now we put only upper bounds on its amplitude \cite{PhysRevD.104.022004}, and on parameters characterizing its directional properties \cite{the_ligo_scientific_collaboration_all-sky_2022}, by the LIGO/Virgo collaboration. For space-based detection, we just have a single detector, cross-correlation based method is not suitable now, so a new SGWB detection algorithm is needed.

SGWBs that are Gaussian, isotropic, and stationary can be fully described by their energy density spectrum \cite{caprini_reconstructing_2019}:
\begin{equation}
	\Omega_{GW}(f)=\frac{f}{\rho_c}\frac{{\rm d}\, \rho_{GW}}{{\rm d}\,f},
\end{equation}
where $\rho_{GW}$ is the energy density of gravitational radiation contained in the frequency range $f$ to $f +df$, $\rho_c = \frac{3H_0^2c^2}{8\pi G}$ is the critical density of the universe, where $c$ is the speed of light, and $G$ is Newton’s constant, and $H_0$ is the Hubble constant. Typical SGWB sources include compact binary coalescence\cite{PhysRevD.104.022004}, cosmic phase transitions \cite{hindmarshGravitationalWavesSound2014}, reheating or preheating after inflation \cite{liuGravitationalWavesOscillons2018}, and primordial scalar and tensor perturbations from inflation \cite{caiGravitationalWavesInduced2019}. There are 2 datasets that contain different types of SGWB signals, TDC-6-1 is generated by the power-law spectrum:
\begin{equation}
	\label{eq:sgwb_pl}
	h^2\Omega_{GW}(f)=10^\alpha\left(\frac{f}{f_*}\right)^{\beta},
\end{equation}
where $h$ is the dimensionless Hubble constant, $f_*$ is the pivot frequency, $\alpha$ characterize its amplitude at $f_*$ and $\beta$ is the slope of the spectrum. The SGWB signal in TDC-6-1 is generated using the parameters $\alpha=-12$ and $\beta = 2/3$.

TDC-6-2 is generated by non-Gaussian scalar perturbation-induced SGWB signals.
The induced GWs are generated during the formation of primordial black holes (PBHs), providing a powerful tool to search for PBHs \cite{yuan_gravitational_2021}.
Typically, curvature perturbations are assumed to have Gaussian probability density functions.
There is a natural expectation that non-Gaussianity will be present when a sharp peak appears in the power spectrum of curvature perturbations.
GWs induced by scalar perturbations with a second-order local-type non-Gaussianity were studied in Ref. \cite{caiGravitationalWavesInduced2019}.
The formulae of the SGWB spectrum used to generate TDC-6-2 data is given in equation (7) in Ref. \cite{caiGravitationalWavesInduced2019} with the parameters $F_{\mathrm{NL}}=10, \sigma=10^{-4}, M_{\mathrm{PBH}}=10^{22}g$, where $F_{\mathrm{NL}}$ characterize the non-Gaussianity, $\sigma$ is the peak width of curvature perturbation spectrum, and $M_{\mathrm{PBH}}$ is the mass of PBH with the assumption that PBHs can serve as all dark matter.

In summary, 9 datasets are designed for TDC, and their essential motivation (i.e. the scientific objective and the technical challenge), and configurations are presented in Table. \ref{tab:Dataset}.
The scientific objectives and the corresponding technical challenges are shown in Tab. \ref{tab:SO} and Tab. \ref{tab:Technical challenge}.

\begin{table}[h!]
	\centering
	\caption{Summary of Taiji Data Challenge datasets}
	\label{tab:Dataset}
	\begin{tabular}{@{} lcccc @{}}
		\toprule
		\textbf{Name} & \textbf{Scientific objective} & \textbf{Technical challenge} & \textbf{Signals}                        & \textbf{Waveform model}        \\
		\midrule
		TDC-1         & 1, 5, 6                       & 7                            & 1 MBHB signal                           & SEOBNRv4\_opt                  \\
		TDC-2-1       & 2, 5, 6                       & 7                            & 1 EMRI signal                           & PN5AAK                         \\
		TDC-2-2       & 2, 5, 6                       & 7                            & 1 EMRI signal                           & XSPEG                          \\
		TDC-3         & 3, 5, 6                       & 7                            & 43 VGB signals                          & Sinusoidal                     \\
		TDC-4         & 3, 5, 6                       & 1, 7                         & $3\times 10^7$ GB signals               & Sinusoidal                     \\
		TDC-5         & 4, 5, 6                       & 2, 7                         & $2\times 10^5$ SOBBH signals            & IMRPhenomD                     \\
		TDC-6-1       & 5, 6, 7                       & 7                            & SGWB                                    & Power-law                      \\
		TDC-6-2       & 5, 6, 7                       & 7                            & SGWB                                    & \begin{tabular}{c}Non-Gaussian \\
			                                                                                                                         curvature       \\
			                                                                                                                         perturbations
		                                                                                                                         \end{tabular} \\
		TDC-7         & 1, 3, 5, 6                    & 1, 3, 7                      & \begin{tabular}{c}10 MBHBs, 43 VGBs, \\
			                                                                               $3\times 10^7\text{ GBs}$
		                                                                               \end{tabular} & \begin{tabular}{c}SEOBNRv4\_opt, \\
			                                                                                               Sinusoidal
		                                                                                               \end{tabular}                         \\
		\bottomrule
	\end{tabular}
\end{table}

\section{Application}
\subsection{File format}

The TDC consists of multiple datasets, with each dataset stored in a separate HDF file., which can be accessed and downloaded from the \href{http://taiji-tdc.ictp-ap.org/dataset_page/}{dataset page} on our TDC website. Within the HDF file, all the data is organized under the attribute \texttt{TDIdata}. Specifically, there is a data array of dimensions \texttt{(4, n\_samples)}, where \texttt{n\_samples} represents the total number of sampling points in the time domain. Each row of the array corresponds to a specific aspect of the data, namely time, TDI X strain, TDI Y strain, and TDI Z strain. This structured arrangement ensures easy and efficient access to the desired information within the dataset.

\subsection{Toy dataset}

\begin{figure}
	\centering
	\includegraphics[width=\textwidth]{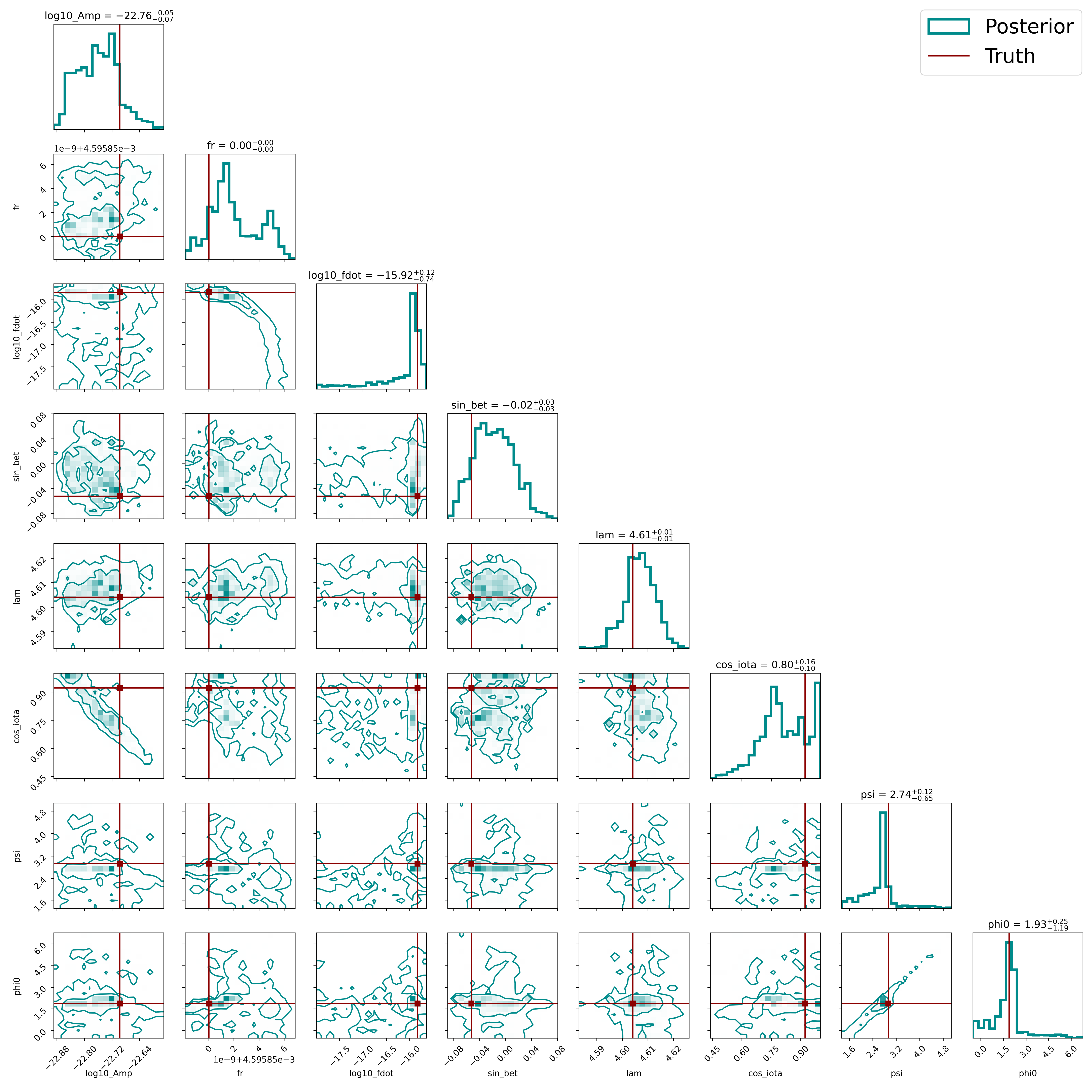}
	\caption{\label{fig:mcmc}MCMC posterior corner plot for the toy dataset. The MCMC posterior analysis is conducted on our toy dataset, utilizing the following source parameter values: $(\mathcal{A}, f, \dot{f}, \lambda, \beta, \iota, \phi_0, \psi) = (2.008192 \times 10^{-23}, 0.00459585, 1.469823 \times 10^{-16}, 4.604082, -0.052144, 0.39811, 1.8777623071795864, 2.927109)$. The posterior distribution is represented by the dark cyan color, while the true parameter values are indicated by the red lines.}
\end{figure}

In this subsection, we present a simple toy dataset, which provides participants with an effective means to validate their pipelines and assess the accuracy of their parameter estimation techniques. The toy dataset has been intentionally designed to be straightforward, featuring only one gravitational wave signal originating from a GB system. By simplifying the dataset to a single signal, participants can precisely evaluate their pipelines without the complexities introduced by waveform models or multiple signals. This simplified setup provides a clear and unambiguous environment for pipeline validation and establishes a solid foundation for accurate parameter estimation. Using this toy dataset, participants have the opportunity to rigorously evaluate the performance of their pipelines. They can assess their pipelines' ability to extract relevant parameters, such as the gravitational wave frequency, amplitude, and arrival time. The well-defined characteristics of the dataset allow participants to directly compare their results, identify any discrepancies, and refine their pipelines accordingly.

To facilitate the effective utilization of the toy dataset, we offer an illustrative example of running a Markov Chain Monte Carlo (MCMC) parameter estimation pipeline. The MCMC pipeline utilized in this demonstration is based on the work by Falxa et al. \cite{falxa_2023}, and the source code can be accessed at the provided link \footnote{ \url{https://gitlab.in2p3.fr/lisa-apc/m3c2}}. This practical demonstration serves as a comprehensive guide, showcasing how the dataset can be leveraged to obtain reliable parameter estimates.

Furthermore, we showcase the outcomes of the parameter estimation process in Fig. \ref{fig:mcmc}, illustrating the attained accuracy and precision through the implementation of the MCMC pipeline on the toy dataset. The visual depiction of the estimated parameters, juxtaposed with the corresponding true values, offers a comprehensive overview of the pipeline's performance and establishes a benchmark for participants to assess and compare their own findings.

By introducing the practical demonstration of the MCMC pipeline and providing visual representations of parameter estimation results, we not only equip participants with the necessary tools to effectively utilize the toy dataset but also prepare them for the real challenge posed by other TDC datasets. The experience gained from working with the toy dataset serves as a valuable cornerstone, allowing participants to build their skills and confidence before tackling the complexities of the actual challenge datasets. The insights and techniques acquired during this process will enable participants to face TDC datasets with increased proficiency and contribute to advancements in space-based gravitational wave data analysis.

\section{Discussions}
Building a data challenge is critical for scientists to develop algorithms and pipelines for future data analysis tasks.
We provide a complete workflow by constructing detection data and waveform templates within and beyond GR.
The datasets and challenges are available on the official website of TDC (\url{http://taiji-tdc.ictp-ap.org}).
Scientists are encouraged to participate in these data challenges and to submit their results.
Currently, there is no deadline, and users are free to submit their results at any time.
We will rank the submitted results for each challenge.
As part of our website, we offer both data downloads and data customization services, which enable users to submit the data configuration file suited to their requirements, and we will provide users links to download the customized dataset.
With TDC, we hope to create a community of researchers who can collaboratively contribute to the development of Taiji's data analysis pipelines and join the journey of exploring the universe and making new discoveries with Taiji Collaboration.

\section{Acknowledgements}
The research was supported by the PengCheng Laboratory Cloud Brain and by PengCheng Cloud-Brain.
Further funding was provided by the National Key Research and Development Program of China Grant No. 2021YFC2203001, No. 2020YFC2201501 \& No. 2021YFC2203002, as well as the NSFC (No. 11920101003, No. 12021003, No. 12173071, No. 12147103, No. 12235019 and No. 12075297).  Z.C and W.H are supported by the CAS Project for Young Scientists in Basic Research YSBR-006. Z.C was also supported by ``The Interdisciplinary Research Funds of Beijing Normal University".

\bibliographystyle{apsrev4-1}
\bibliography{references}

\end{document}